\documentclass[a4paper,11pt]{article}
\usepackage[dvips]{color,graphicx}

\def\EN{\hfill\rule{2 truemm}{3 truemm}}

\newcommand{\vs}[1]{\vspace{#1 mm}}

\renewcommand{\d}{\delta}

\newcommand{\pa}{\partial}

\usepackage{graphicx,amssymb,amsmath,bm,latexsym}

\begin{document}

\begin{titlepage}

\vskip .5in

\begin{center}

{\Large\bf On $W_{1+\infty}$  3-algebra and integrable system } \vskip .5in

{\large Min-Ru Chen$^{a,b,}$\footnote{cmr@henu.edu.cn},
Shi-Kun Wang$^{c,}$\footnote{wsk@amss.ac.cn},
Xiao-Li Wang$^{a,d,}$\footnote{wxldllg$_{-}$2001@163.com}
Ke Wu$^{a,e,}$\footnote{wuke@mail.cnu.edu.cn}\\
and Wei-Zhong Zhao$^{a,e,f,}$
\footnote{Corresponding author: zhaowz100@163.com}} \\
\vs{10}
$^a${\em School of Mathematical Sciences, Capital Normal University,
Beijing 100048, China} \\
$^b${\em College of Mathematics and Information Sciences, Henan
University, Kaifeng 475001, China} \\
$^c${\em Institute of Applied Mathematics, Academy of Mathematics and Systems Science, Chinese
Academy of Sciences, Beijing 100190, China} \\
$^d${\em School of Science, Qilu University of Technology, Jinan 250353, China }\\
$^e${\em Beijing Center for Mathematics and Information
Interdisciplinary Sciences, Beijing 100048, China } \\
$^f${\em Institute of Mathematics and Interdisciplinary Science,
Capital Normal University, Beijing 100048, China } \\

\vskip .2in \vspace{.3in}
\begin{abstract}
We construct the $W_{1+\infty}$  3-algebra and investigate the relation between this
infinite-dimensional 3-algebra and the integrable systems. Since the $W_{1+\infty}$
3-algebra with a fixed generator $W^0_0$ in the operator Nambu 3-bracket
recovers  the $W_{1+\infty}$ algebra, it is natural to derive the  KP hierarchy
from the Nambu-Poisson evolution equation. For the general case of the $W_{1+\infty}$
3-algebra, we directly derive the KP and KdV equations from the Nambu-Poisson evolution
equation with the different Hamiltonian pairs. We also discuss the connection between
the $W_{1+\infty}$ 3-algebra and the dispersionless KdV equations. Due to the
Nambu-Poisson evolution equation involves two Hamiltonians, the deep relationship
between the Hamiltonian pairs of KP hierarchy is revealed.
Furthermore we give a realization of  $W_{1+\infty}$ 3-algebra  in terms of
a complex bosonic field. Based on the Nambu 3-brackets of the complex bosonic field, we
derive the (generalized) nonlinear Schr\"{o}dinger equation and give an application
in optical soliton.

\end{abstract}

\end{center}



\vfill

\end{titlepage}

%%%%%%%%%%%%%%%%%%%%%%%%%%%%%%%%%%%%%%%%%%%%%%%%%%%%%%%%%%%
\section{Introduction}
%%%%%%%%%%%%%%%%%%%%%%%%%%%%%%%%%%%%%%%%%%%%%%%%%%%%%%%%%%%

The infinite-dimensional algebras  play a very important role in the study of
integrable systems. This is due to the fact that they have a deep intrinsic
connection to the integrable systems. The Virasoro algebra is an important
infinite-dimensional algebra which has been turned out to be
a universal symmetry algebra in two-dimensional conformal field theory.
Gervais and Neveu \cite{Gervais1, Gervais2} found that this infinite-dimensional algebra
is  intrinsically related to the Korteweg-de Vries (KdV) equation where the Poisson bracket
of the KdV equation is indeed the classical Virasoro algebra.
The lattice Virasoro algebra
appears in the study of the Toda field theory and Toda integrable systems \cite{Faddeev, Inoue}.
Furthermore the intrinsic connection between  the super Virasoro algebra and super KdV equation
has been studied in \cite{Kupershmidt}-\cite{Tanaka2}.
As a higher-spin extension of the
Virasoro algebra, the $W_{N}$ algebra has been found to be
associated to the generalized KdV hierarchies \cite{Mathieu2, Yamagishi}.

The KP hierarchy is an integrable system consisting of an infinite
number of non-linear differential equations.
This hierarchy has proven to be bi-Hamiltonian and to
possess an infinite number of  conserved quantities.
All the (generalized) KdV hierarchies can be incorporated into the KP hierarchy.
The  $W_{1+\infty}$ algebra \cite{PRS}
can be regards as  a linear deformation of the
$W_N$ algebras in the large N limit \cite{Bakas}.
It has very important applications in
physics, such as quantum Hall effect \cite{Cappelli},
topological string \cite{Dijkgraaf}, Alday-Gaiotto-Tachikawa (AGT) relation \cite{Kanno},
2-dimensional quantum gravity \cite{Itoyama} and crystal melting \cite{Vafa}.
In the context of integrable systems,
it was shown that the  $W_{1+\infty}$ algebra
is  intrinsically related  to the first Hamiltonian structure of the KP
hierarchy \cite{Yu, Yamagishi1991}.
The second Hamiltonian structure of the KP hierarchy  was then shown
to be isomorphic to $\hat W_{\infty}$ \cite{Figueroa}, a centerless deformation of $W_{\infty}$.

Nambu mechanics \cite{Nambu}
is a generalization of classical Hamiltonian mechanics in which
the usual binary Poisson bracket is replaced by the  Nambu bracket.
Based on the Nambu bracket,
a notion of a Nambu 3-algebra was introduced in
\cite{Takhtajan}  as a natural generalization of a Lie algebra for
higher-order algebraic operations.
Recently 3-algebras have been paid great attention due to a world-volume description of multiple M2-branes
proposed by Bagger and Lambert \cite{BL2007}, and Gustavsson \cite{Gustavsson} (BLG).
There has been an increasing interest in the applications of 3-algebras to the string/M-theory
\cite{CS2008}-\cite{Chenho}.
The properties of various 3-algebras especially the infinite-dimensional cases
have been widely investigated,  such as Virasoro-Witt 3-algebra \cite{Curtright}-\cite{Larsson},
Kac-Moody 3-algebra \cite{Lin} and  $w_{\infty}$ 3-algebra
\cite{Chakrabortty, Chen}.  Furthermore the q-deformation of  Virasoro-Witt 3-algebra
has also been investigated.  The nontrivial q-deformed Virasoro-Witt 3-algebra constructed in \cite{Ding}
is the so-called sh-3-Lie algebra.

More recently, the relation between the infinite-dimensional
3-algebras and the integrable systems has been investigated.
Chen er al. \cite{Chen2012} investigated the  classical Heisenberg and $w_{\infty}$
3-algebras and established the relation between the dispersionless KdV hierarchy and these two
infinite-dimensional 3-algebras. They found that  the dispersionless KdV system is
not only a bi-Hamiltonian system, but also a  bi-Nambu-Hamiltonian system.
It should be pointed out that the dispersionless KdV hierarchy is a very simple integrable system.
An interesting open question is whether we can derive other integrable nonlinear evolution equations
from the infinite-dimensional 3-algebras.
We have mentioned that the first Hamiltonian structure of the KP
hierarchy  is indeed corresponding to the $W_{1+\infty}$ algebra.
In this paper, we shall construct the $W_{1+\infty}$  3-algebra and discuss the relation
between this infinite-dimensional 3-algebra and the integrable systems.

This paper is organized as follows. In section 2, we construct the $W_{1+\infty}$ 3-algebra.
In section 3, we establish the relation between the $W_{1+\infty}$ 3-algebra
and the KP hierarchy in the framework of Nambu mechanics.  In section 4, By means of the
$W_{1+\infty}$ 3-algebra, We derive the KP and KdV equations from the Nambu-Poisson (N-P) evolution
equation with the different Hamiltonian pairs. In section 5, we establish the connections
between the $W_{1+\infty}$ 3-algebra and the dispersionless KdV equations.
A conjecture is presented.  In section 6, we give a realization of  $W_{1+\infty}$ 3-algebra
and investigate the (generalized) nonlinear Schr\"{o}dinger equation.  Moreover we give
an application of the generalized nonlinear Schr\"{o}dinger equation derived from the N-P evolution
equation. We end this paper with the
concluding remarks in section 7.

%%%%%%%%%%%%%%%%%%%%%%%%%%%%%%%%%%%%%%%%%%%%%%%%%%%%%%%%%%%
\section{ $W_{1+\infty}$ 3-algebra}
%%%%%%%%%%%%%%%%%%%%%%%%%%%%%%%%%%%%%%%%%%%%%%%%%%%%%%%%%%%
%%%%%%%%%%%%%%%%%%%%%%%%%%%%%%%%%%%%%%%%%%%%%%%%%%%%%%%%%%%
\subsection{ $W_{1+\infty}$ algebra}
%%%%%%%%%%%%%%%%%%%%%%%%%%%%%%%%%%%%%%%%%%%%%%%%%%%%%%%%%%%

Let us take  $\{W_m^r=(-1)^r e^{{\bf i}mx}\frac{d^r}{dx^r}|m,r\in \mathbb{Z}, r\geq 0, {\bf i}=\sqrt {-1}\}$.
We then obtain the algebra
\begin{eqnarray}\label{eq:walgebra}
[W_m^r, W_n^s]
&=&(\sum_{p=0}^r C_r^p (-{\bf i}n)^p-\sum_{p=0}^s C_s^p (-{\bf i}m)^p)W_{m+n}^{r+s-p},
\end{eqnarray}
where  $C_r^p=\frac{r(r-1)\cdots(r-p+1)}{p!}$.
The first
few brackets read
\begin{eqnarray}\label{eq:walgebra'}
&&[W_m^0, W_n^0]=0,\nonumber\\
&&[W_m^0, W_n^1]={\bf i}mW_{m+n}^{0},\nonumber\\
&&[W_m^1, W_n^1]={\bf i}(m-n)W_{m+n}^{1},\nonumber\\
&&[W_m^0, W_n^2]=2{\bf i}mW_{m+n}^{1}+m^2W_{m+n}^{0},\nonumber\\
&&[W_m^0, W_n^3]=3{\bf i}mW_{m+n}^{2}+3m^2W_{m+n}^{1}-{\bf i}m^3W_{m+n}^0,\nonumber\\
&&[W_m^1, W_n^2]={\bf i}(2m-n)W_{m+n}^{2}+m^2W_{m+n}^{1},\nonumber\\
&&[W_m^2, W_n^2]={\bf i}(2m-2n)W_{m+n}^{3}+(m^2-n^2)W_{m+n}^{2},\nonumber\\
&&[W_m^2, W_n^3]={\bf i}(3m-2n)W_{m+n}^{4}+(3m^2-n^2)W_{m+n}^{3}-{\bf i}m^3W_{m+n}^{2}.
\end{eqnarray}
The first three commutators can be recognized as the semidirect product of the Witt algebra
with an abelian current algebra.

Taking the Fourier transformation field (FTF)
\begin{eqnarray}\label{eq:FTF1}
v_r(x)=\frac{-{\bf i}}{2\pi}\sum_{m=-\infty}^{\infty} W_m^r e^{-{\bf i}mx},
\end{eqnarray}
and defining $\{,\}=-{\bf i}[,]$, we may rewrite (\ref{eq:walgebra}) as
the following Poisson bracket algebra derived in \cite{Yu}:
\begin{eqnarray}\label{eq:fwalgebra}
&&\{v_r(x), v_s(y)\}
=\sum_{p=0}^s C_s^p (-1)^p v_{r+s-p}(y)\partial_y^p\delta(x-y)-\sum_{p=0}^r C_r^p (-1)^p v_{r+s-p}(x)\partial_x^p\delta(x-y)\nonumber\\
&&=\sum_{p=0}^s\sum_{l=0}^p C_s^p C_p^l v_{r+s-p}^{(l)}(x)(\partial_x^{p-l}\delta(x-y))-\sum_{p=0}^r C_r^p (-1)^p v_{r+s-p}(x)\partial_x^p\delta(x-y).
\end{eqnarray}
The $W_{1+\infty}$ algebra is an infinite-dimensional
higher-spin algebra with generators having all integer conformal spins $s\geq 1$ \cite{PRS}.
It is worth noting that the Poisson bracket algebra (\ref{eq:fwalgebra}) has been proved to be
isomorphic to the
$W_{1+\infty}$ algebra \cite{Yu}. Thus we call (\ref{eq:walgebra}) the $W_{1+\infty}$ algebra.

Another important infinite-dimensional algebra which should be mentioned is the higher-order Virasoro algebra (HOVA).
By investigating the higher-order differential neighbourhoods
in holomorphic mappings of $S^1$ a vector space, Zha \cite{Zha} first constructed this infinite-dimensional algebra.
Its generators are $\{L_m^r=(-1)^r z^{m+r}\frac{d^r}{dz^r}|m,r\in\mathbb{Z}, r\geq 0\}$.
Taking $z=e^{{\bf i}x}$, we find that the generators of HOVA
and $W_{1+\infty}$  algebra can be linearly expressed reciprocally. It implies  that the $W_{1+\infty}$ algebra (\ref{eq:walgebra})  is also indeed isomorphic to the
HOVA.

%%%%%%%%%%%%%%%%%%%%%%%%%%%%%%%%%%%%%%%%%%%%%%%%%%%%%%%%%%%
\subsection{ $W_{1+\infty}$ 3-algebra}
%%%%%%%%%%%%%%%%%%%%%%%%%%%%%%%%%%%%%%%%%%%%%%%%%%%%%%%%%%%

The operator Nambu 3-bracket was  defined
to be a sum of commutators of two operators multiplying
the remaining one \cite{Nambu}, i.e.,
\begin{eqnarray}\label{eq:3bracket}
[A, B, C]=[A, B]C+[B, C]A+[C, A]B,
\end{eqnarray}
where $[A, B]=AB-BA$.

Based on the operator Nambu 3-bracket (\ref{eq:3bracket})
and the commutation relation (\ref{eq:walgebra}),
after a straightforward calculation
we obtain the $W_{1+\infty}$ 3-algebra
\begin{eqnarray}\label{eq:3-alg1}
&&[W_m^r, W_n^s, W_k^h]
=(\sum_{p=0}^r (-{\bf i})^p C_r^p n^p-\sum_{p=0}^s (-{\bf i})^p C_s^p m^p)\sum_{q=0}^{r+s-p}
(-{\bf i})^q C_{r+s-p}^qk^q W_{m+n+k}^{r+s+h-p-q}\nonumber\\
&&\qquad+(\sum_{p=0}^s(-{\bf i})^p C_s^p k^p-\sum_{p=0}^h(-{\bf i})^p C_h^p n^p)\sum_{q=0}^{s+h-p}
(-{\bf i})^q C_{s+h-p}^qm^q W_{m+n+k}^{r+s+h-p-q}\nonumber\\
&&\qquad+(\sum_{p=0}^h(-{\bf i})^p C_h^p m^p-\sum_{p=0}^r(-{\bf i})^p C_r^p k^p)\sum_{q=0}^{h+r-p}
(-{\bf i})^q C_{h+r-p}^qn^q W_{m+n+k}^{r+s+h-p-q}.
\end{eqnarray}

Taking $k=h=0$ in (\ref{eq:3-alg1}), we note that
the Nambu 3-bracket $[W_m^r, W_n^s, W_0^0]$ can be regarded as the
parametrized bracket relation $[W_m^r, W_n^s]_{W_0^0}$ which satisfies the Jacobi
identity.  This parametrized bracket relation gives rise to the  $W_{1+\infty}$ algebra (\ref{eq:walgebra})
\begin{eqnarray}\label{eq:www00}
[W_m^r, W_n^s]_{W_0^0}
\equiv[W_m^r, W_n^s, W_0^0]
=[W_m^r, W_n^s].
\end{eqnarray}

It is known that all three types of 3-brackets do satisfy  a 3-on-3-on-3 multiple bracket relation, i.e., so-called
BI condition \cite{Bremner, Curtright2009}
\begin{eqnarray}\label{eq:BI}
\epsilon^{i_1i_2\cdots i_6}[[A, [B_{i_1}, B_{i_2}, B_{i_3}], B_{i_4}], B_{i_5}, B_{i_6}]=\epsilon^{i_1i_2\cdots i_6}[[A, B_{i_1}, B_{i_2}], [B_{i_3}, B_{i_4}, B_{i_5}], B_{i_6}] ,
\end{eqnarray}
where $i_1, \cdots, i_6$ are implicitly summed from 1 to 6. Thus the BI may be confirmed to hold for the ternary algebra
(\ref{eq:3-alg1}).

It is worthwhile to mention another condition, i.e.,  Filippov condition or fundamental identity (FI) condition \cite{Filippov}
\begin{eqnarray}\label{eq:FI}
[A, B, [C, D, E]]=[[A, B, C], D, E]+[C, [A, B, D], E]+[C, D, [A, B, E]].
\end{eqnarray}
The FI condition (\ref{eq:FI}) is not an operator identity. It holds only in special circumstances.
For the $W_{1+\infty}$ 3-algebra (\ref{eq:3-alg1}),  we find that it does not satisfy the
FI condition (\ref{eq:FI}).

In order to achieve a better understanding of the $W_{1+\infty}$ 3-algebra (\ref{eq:3-alg1}),
we list the first few brackets as follows:
\begin{eqnarray*}
\ [W_m^0, W_n^0, W_k^0]&=&0,\nonumber\\
\ [W_m^1, W_n^0, W_k^0]&=&{\bf i}(k-n)W_{m+n+k}^0,\nonumber\\
\ [W_m^1, W_n^1, W_k^0]&=&{\bf i}(m-n)W_{m+n+k}^1+(km-kn)W_{m+n+k}^0,\nonumber\\
\ [W_m^1, W_n^1, W_k^1]&=&0,\nonumber\\
\ [W_m^2, W_n^0, W_k^0]&=&{\bf i}(2k-2n)W_{m+n+k}^1+(k^2-n^2)W_{m+n+k}^0,\nonumber\\
\ [W_m^3, W_n^0, W_k^0]&=&{\bf i}(3k-3n)W_{m+n+k}^2+(3k^2-3n^2)W_{m+n+k}^1+i(n^3-k^3)W_{m+n+k}^0,\nonumber\\
\ [W_m^1, W_n^2, W_k^0]&=&{\bf i}(2m-n-k)W_{m+n+k}^2+(m^2-k^2+2km-2kn)W_{m+n+k}^1\nonumber\\
&&+{\bf i}(nk^2-2k^2m-km^2)W_{m+n+k}^0,\nonumber\\
\ [W_m^2, W_n^2, W_k^0]&=&{\bf i}(2m-2n)W_{m+n+k}^3+(m^2-n^2+4km-4kn)W_{m+n+k}^2\nonumber\\
&&+{\bf i}(2kn^2-2km^2+6nk^2-6mk^2)W_{m+n+k}^1\nonumber\\
&&+(k^2n^2-k^2m^2+2nk^3-2k^3m)W_{m+n+k}^0,\nonumber\\
\ [W_m^1, W_n^1, W_k^2]&=&{\bf i}(n-m)W_{m+n+k}^{3}+(n^2-m^2-kn-km)W_{m+n+k}^{3}\nonumber\\
&&+{\bf i}(km^2-kn^2-m^2n+mn^2)W_{m+n+k}^1,\nonumber\\
\end{eqnarray*}
\begin{eqnarray}\label{eq:3-alg2}
\ [W_m^1, W_n^2, W_k^2]&=&{\bf i}(n-k)W_{m+n+k}^{4}+(n^2-k^2+2mn-2mk)W_{m+n+k}^{3}\nonumber\\
&&+{\bf i}(5km^2-5nm^2+k^2-kn^2)W_{m+n+k}^{2}\nonumber\\
&&+(k^2m^2-m^2n^2+2km^3-2m^3n)W_{m+n+k}^{1},\nonumber\\
\ [W_m^2, W_n^2, W_k^2]&=&2(k^3n-k^3m+m^3k-m^3n+n^3m-n^3k)W_{m+n+k}^{2}\nonumber\\
&&+4{\bf i}(k^2n-k^2m+m^2k-m^2n+n^2m-n^2k)W_{m+n+k}^{1}.
\end{eqnarray}
We observe that these brackets do not close except for the generators $W_m^i, i=0,1$.
The  Nambu 3-bracket (\ref{eq:3bracket}) with respect to the generators $W_m^0$ and $W_n^1$ indeed give the so-called
the Virasoro-Witt 3-algebra \cite{Curtright, Larsson}.  However, this 3-algebra  does not satisfy the FI condition (\ref{eq:FI}).
The general Virasoro-Witt 3-algebra is given by Curtright et al. \cite{Curtright}
 \begin{eqnarray}\label{eq:V-W-3-alg}
\ [R_p, R_q, R_k]&=&0,\nonumber\\
\ [Q_p, R_q, R_k]&=&(k-q)R_{k+p+q},\nonumber\\
\ [Q_p, Q_q, R_k]&=&(p-q)(Q_{k+p+q}+zkR_{k+p+q}),\nonumber\\
\ [Q_k, Q_m, Q_n]&=&(k-m)(m-n)(k-n)R_{k+m+n},
\end{eqnarray}
where $z$ is a parameter.
They showed that the Virasoro-Witt 3-algebra (\ref{eq:V-W-3-alg}) does not satisfy
the FI condition (\ref{eq:FI}), except when $z = \pm 2{\bf i}$.

In order to derive the classical $w_{1+\infty}$ 3-algebra,
we take the generators to be
$\{W_m^r=(-{\bf i}\hbar)^r e^{{\bf i}mx}\frac{d^r}{dx^r}|m,r\in \mathbb{Z}, r\geq 0\}$,
where $\hbar$ is introduced into the generators.
By means of (\ref{eq:3-alg1}) and classical limit relation
$\{, , \}=\lim_{\hbar\rightarrow 0}\frac{1}{{\bf i}\hbar}[, ,]$,
we have the classical $w_{1+\infty}$ 3-algebra as follows:
\begin{eqnarray}\label{eq:cw3alg}
\{W_m^r,W_n^s,W_k^h\}&=&\lim_{\hbar\rightarrow 0}\frac{1}{{\bf i}\hbar}[W_m^r,W_n^s,W_k^h]\nonumber\\
&=&{\bf i}(r(k-n)+s(m-k)+h(n-m))W_{m+n+k}^{r+s+h-1}.
\end{eqnarray}
Not as the case of $W_{1+\infty}$ 3-algebra, we find that the classical
$w_{1+\infty}$ 3-algebra (\ref{eq:cw3alg}) satisfies the  FI condition.

%%%%%%%%%%%%%%%%%%%%%%%%%%%%%%%%%%%%%%%%%%%%%%%%%%%%%%%%%%%
\section{  $W_{1+\infty}$ 3-algebra and KP hierarchy}
%%%%%%%%%%%%%%%%%%%%%%%%%%%%%%%%%%%%%%%%%%%%%%%%%%%%%%%%%%%
The KP hierarchy is a paradigm of the hierarchies of integrable systems.
It is defined as an infinite system of equations given in Lax form by
\begin{eqnarray}\label{eq:KPh}
\frac{\partial L}{\partial t_n}=[B_n, L],\quad n=1,2,\cdots,
\end{eqnarray}
under the compatible condition, (\ref{eq:KPh}) is equivalent to the Zakharov-Shabat (ZS) equation
\begin{eqnarray}\label{eq:KPhe}
\frac{\partial B_m}{\partial t_n}-\frac{\partial B_n}{\partial t_m}+[B_m, B_n]=0,
\end{eqnarray}
where the pseudo-differential operator $L$ is given by
\begin{eqnarray}\label{eq:L}
L=\partial+\sum_{i=0}^{+\infty}v_i\partial^{-i-1},
\end{eqnarray}
and $B_n=(L)^n_+$ is the differential part of $L^n$.

The Hamiltonians of the KP Hierarchy are given by the general expression
\begin{eqnarray}\label{eq:KPGHamiltonian}
H_n=Tr\frac{L^{n+1}}{n+1}=\frac{1}{n+1}\int Res L^{n+1}dx.
\end{eqnarray}
The first few members are
\begin{eqnarray}\label{eq:KPHamiltonian}
H_0&=&\int v_0dx,\nonumber\\
H_1&=&\int v_1dx,\nonumber\\
H_2&=&\int (v_2+v_0^2)dx,\nonumber\\
H_3&=&\int (v_3+3v_0v_1)dx,\nonumber\\
H_4&=&\int (v_4+4v_0v_2+2v_1^2+2v_0^3+2 v_0v_{1,x}+v_0v_{0,xx})dx,\nonumber\\
H_5&=&\int (v_5+5v_0v_3+5v_1v_2+10v_1v_0^2-5 v_2v_{0,x}+4 v_1v_{0,xx}- v_0v_{0,xxx}+ v_0v_{1,xx})dx,\nonumber\\
H_6&=&\int (v_6+6v_0v_4+6v_1v_3+3v_2^2+5 v_0^4+15 v_0^2v_2+9v_0v_{3,x}+3v_1v_{2,x}+15 v_0v_1^2\nonumber\\
&&+15 v_0v_{0,x}v_1+13v_0v_{2,xx}+2v_{0,x}v_{2,x}+4 v_1v_{1,xx}+7v_0v_{1,xxx}+3v_1v_{0,xxx}\nonumber\\
&&+15v_0^2v_{1,x}+5v_0^2v_{0,xx}+v_{0,xx}^2)dx,\nonumber\\
H_7&=&\int (v_7+7v_0v_5+7v_1v_4+7v_2v_3+21 v_0^2v_3+7v_1^3+42 v_0v_1v_2+14v_0v_{4,x}+7v_1v_{3,x}+7v_0v_{3,xx}\nonumber\\
&&+14 v_{0,xx}v_3+14 v_1v_{2,xx}+\frac{77}{4}v_0v_{2,xxx}+\frac{21}{4}v_2v_{0,xxx}+\frac{7}{4}v_{0,xxx}v_{1,x}+\frac{35}{4}v_0v_{1,xxxx}\nonumber\\
&&+42v_0v_{0,xx}v_1+\frac{7}{2}v_0^2v_{1,xx}+21v_0^2v_{2,x}+21v_0v_1v_{1,x}+21v_{0,x}^2v_1+35v_0^3v_1)dx,\nonumber\\
&&\vdots,
\end{eqnarray}
where the subscript $x$ denotes the derivative with respect to the variable $x$.

It is known that
the Poisson bracket (\ref{eq:fwalgebra}) leads to the Hamiltonian description of
the KP flow equations  \cite{Yu, Yamagishi1991}.
Substituting the  Hamiltonians (\ref{eq:KPGHamiltonian}) into
the Poisson evolution equation
\begin{eqnarray}\label{eq:Pee}
\frac{\partial v_i(t, x)}{\partial t_n}=\{v_i, H_n\},
\end{eqnarray}
and using the Poisson bracket (\ref{eq:fwalgebra}),
we may derive the KP hierarchy.

In order to establish the relation between the $W_{1+\infty}$ 3-algebra and integrable systems,
let us introduce the FTF as follows:
\begin{eqnarray}\label{eq:FTF2}
v_r(x)=\frac{1+{\bf i}}{2\sqrt{2}\pi}\sum_{m=-\infty}^{\infty} W_m^r e^{-{\bf i}mx},
\end{eqnarray}
and define the classical Nambu bracket $\{\ ,\ , \}=-{\bf i}[\ , \ , \ ]$.
Then we can rewrite the $W_{1+\infty}$
3-algebra (\ref{eq:3-alg1}) as the following Nambu 3-bracket relation:
\begin{eqnarray}\label{eq:KP3-alg2}
&&\{v_r(x), v_s(y), v_h(z)\}=\nonumber\\
&&\sum_{p=0}^r\sum_{q=0}^{r+s-p} (-1)^{p+q}C_r^p C_{r+s-p}^q v_{r+s+h-p-q}(x)
\partial_x^p\delta(x-y) \partial_x^q\delta(x-z)\nonumber\\
&&-\sum_{p=0}^s\sum_{q=0}^{r+s-p} (-1)^{p+q}C_s^p C_{r+s-p}^q v_{r+s+h-p-q}(y)
\partial_y^p\delta(y-x) \partial_y^q\delta(y-z)\nonumber\\
&&+\sum_{p=0}^s\sum_{q=0}^{s+h-p} (-1)^{p+q}C_s^p C_{s+h-p}^q v_{r+s+h-p-q}(y)
\partial_y^p\delta(y-z) \partial_y^q\delta(y-x)\nonumber\\
&&-\sum_{p=0}^h\sum_{q=0}^{s+h-p} (-1)^{p+q}C_h^p C_{s+h-p}^q v_{r+s+h-p-q}(z)
\partial_z^p\delta(z-y) \partial_z^q\delta(z-x)\nonumber\\
&&+\sum_{p=0}^h\sum_{q=0}^{h+r-p} (-1)^{p+q}C_h^p C_{h+r-p}^q v_{r+s+h-p-q}(z)
\partial_z^p\delta(z-x) \partial_z^q\delta(z-y)\nonumber\\
&&-\sum_{p=0}^r\sum_{q=0}^{h+r-p} (-1)^{p+q}C_r^p C_{h+r-p}^q v_{r+s+h-p-q}(x)
\partial_x^p\delta(x-z) \partial_x^q\delta(x-y).
\end{eqnarray}

Not as the Poisson evolution equation, the N-P evolution equation involves
two Hamiltonians.  It is given by
\begin{eqnarray}\label{eq:Nee}
\frac{\partial v_i}{\partial t_{mn}}=\{v_i, H_m, H_n\}.
\end{eqnarray}

In terms of the Nambu 3-bracket relation (\ref{eq:KP3-alg2}), we have
\begin{eqnarray}\label{eq:vH0v}
\{v_r, H_0, v_s\}=\sum_{p=0}^s C_s^p (-1)^p v_{r+s-p}(y)\partial_y^p\delta(x-y)
-\sum_{p=0}^r C_r^p (-1)^p v_{r+s-p}(x)\partial_x^p\delta(x-y).
\end{eqnarray}
Comparing (\ref{eq:vH0v}) with (\ref{eq:fwalgebra}), it gives the equivalent relation
\begin{eqnarray}\label{eq:vHv}
\{v_r, H_0, v_s\}\cong\{v_r, v_s\},
\end{eqnarray}
where the FTFs in Nambu 3-bracket and Poisson bracket are given by
(\ref{eq:FTF2}) and (\ref{eq:FTF1}), respectively.
This equivalent relation is indeed corresponding to the relation (\ref{eq:www00}) with
$\{\ ,\ , \}=-{\bf i}[\ , \ , \ ]$ and $\{\ , \}=-{\bf i}[\ , \ ]$.
Due to (\ref{eq:www00}), it can be  verified that the triple Hamiltonians $(H_0, H_n, H_m)$
satisfy $\{H_0,  H_n,  H_m\} = 0$.
But for the triple Hamiltonians $(H_l, H_n, H_m)$, $l,n,m\in Z_{+}$,
by means of (\ref{eq:KP3-alg2}), we note that they are not in involution with the
Nambu 3-bracket structure, i.e., $\{H_l, H_n, H_m\}\neq 0$.

For the Hamiltonian pairs $(H_0, H_n)$, by means of (\ref{eq:vHv}), it is not difficult to
show that
\begin{eqnarray}\label{eq:HPR}
\{v_i, H_0, H_n\}\cong\{v_i, H_n\}.
\end{eqnarray}
Thus based on (\ref{eq:KP3-alg2}) and (\ref{eq:HPR}),
it is natural to derive the KP hierarchy from the N-P evolution equation (\ref{eq:Nee}) with $H_m=H_0$.
For later convenience, we list the first few members of the hierarchy here.

$\bullet$ Let us consider the following equations:
\begin{subequations}\label{eq:KP1}
\begin{align}
&&\frac{\partial v_0}{\partial t_{03}}=\{v_0, H_0,  H_3\}=3 v_{2,x}+3 v_{1,xx}+ v_{0,xxx}+6 v_0v_{0,x},\label{eq:KP1a}\\
&&\frac{\partial v_0}{\partial t_{02}}=\{v_0, H_0, H_2\}=2 v_{1,x}+ v_{0,xx},\ \ \ \ \ \ \ \ \ \ \ \ \ \ \ \ \ \ \ \ \ \ \ \ \ \label{eq:KP1b}\\
&&\frac{\partial v_1}{\partial t_{02}}=\{v_1, H_0, H_2\}=2 v_{2,x}+ v_{1,xx}+2 v_0v_{0,x}.\ \ \ \ \ \ \ \ \ \ \ \ \label{eq:KP1c}
\end{align}
\end{subequations}
From (\ref{eq:KP1b}) and (\ref{eq:KP1c}), we obtain
\begin{eqnarray}\label{eq:kpv1-4}
v_1&=&\frac{1}{2}(\partial_x^{-1}v_{0,t_{02}}-v_{0,x}),\nonumber\\
v_2&=&\frac{1}{4}(\partial_x^{-2}v_{0,t_{02}t_{02}}-2v_{0,t_{02}}+v_{0,xx}-2 v_0^2).
\end{eqnarray}
Substituting (\ref{eq:kpv1-4}) into (\ref{eq:KP1a}),
we obtain the usual KP equation
\begin{eqnarray}\label{eq:KPE}
3v_{0,yy}=(4v_{0,t}-v_{0,xxx}-12v_0v_{0,x})_x,
\end{eqnarray}
where $y=t_{02}$ and $t=t_{03}$.
The KP hierarchy can also be obtained from (\ref{eq:KPhe}).
It is easy to derive the KP equation (\ref{eq:KPE})
from (\ref{eq:KPhe}) by taking $m=2$ and $n=3$.

$\bullet$ To give the integrable equation corresponding to the case of $m=2$ and $n=4$ in (\ref{eq:KPhe}), we
consider the following equations:
\begin{subequations}\label{eq:KP2}
\begin{align}
&&\frac{\partial v_0}{\partial t_{04}}=\{v_0, H_0, H_4\}=4 v_{3,x}+6 v_{2,xx}
+4 v_{1,xxx}+ v_{0,xxxx}\ \nonumber\\
&&+6 (v_{0,x})^2+6 v_0v_{0,xx}+12 v_0v_{1,x}+12 v_1v_{0,x},\label{eq:KP2a}\\
&&\frac{\partial v_0}{\partial t_{02}}=\{v_0, H_0,  H_2\}=2 v_{1,x}+ v_{0,xx},\ \ \ \ \ \ \ \ \ \ \ \ \ \ \ \ \ \ \ \ \ \ \ \ \ \ \  \label{eq:KP2b}\\
&&\frac{\partial v_1}{\partial t_{02}}=\{v_1, H_0,  H_2\}=2 v_{2,x}+ v_{1,xx}+2 v_0v_{0,x},\ \ \ \ \ \ \ \ \ \ \ \ \ \ \label{eq:KP2c}\\
&&\frac{\partial v_2}{\partial t_{02}}=\{v_2, H_0, H_2\}=2 v_{3,x}+ v_{2,xx}-2 v_0v_{0,xx}+4 v_1v_{0,x}.\label{eq:KP2d}
\end{align}
\end{subequations}
From (\ref{eq:KP2b}), (\ref{eq:KP2c}) and (\ref{eq:KP2d})  we may give the expressions of
$v_1$, $v_2$ and $v_3$ with respect to $v_0$, respectively. Then substituting them
 into (\ref{eq:KP2a}),  we obtain
\begin{eqnarray}\label{eq:ekp2}
\frac{\partial^3 v_0}{\partial y\partial x^2}-2\frac{\partial v_0}{\partial t}
+\partial_x^{-2}\frac{\partial^3 v_0}{\partial y^3}+4 v_{0,x}\partial_x^{-1}\frac{\partial v_0}{\partial y}+8v_0\frac{\partial v_0}{\partial y}=0,
\end{eqnarray}
where $y=t_{02}$ and $t=t_{04}$.

$\bullet$ The integrable equation corresponding to the case of $m=2$ and $n=5$ in (\ref{eq:KPhe})
can be derived from the following equations:
\begin{eqnarray}\label{eq:KP3}
&&\frac{\partial v_0}{\partial t_{05}}=\{v_0, H_0, H_5\}, \ \
\frac{\partial v_0}{\partial t_{02}}=\{v_0, H_0, H_2\},\ \
\frac{\partial v_1}{\partial t_{02}}=\{v_1, H_0, H_2\},\nonumber\\
&&\frac{\partial v_2}{\partial t_{02}}=\{v_2, H_0, H_2\},\ \
\frac{\partial v_3}{\partial t_{02}}=\{v_3, H_0, H_2\}.
\end{eqnarray}
The integrable equation is
\begin{eqnarray}\label{eq:ekp3}
&&\frac{\partial v_0}{\partial t}=\frac{5}{16}\partial_x^{-3}v_{0,yyyy}+\frac{5}{8}v_{0,xyy}+\frac{5}{2}v_{0,y}\partial_x^{-1}v_{0,y}
+\frac{5}{2}v_{0,x}v_{0,xx}+\frac{5}{4}v_{0}v_{0,xxx}+\frac{1}{16}v_{0,xxxxx}\nonumber\\
&&\qquad\quad +\frac{5}{2}v_{0}\partial_x^{-1}v_{0,yy}+\frac{5}{4}\partial_x^{-1}(v_{0,y})^2+\frac{5}{4}\partial_x^{-1}(v_0v_{0,yy})
+\frac{5}{4}v_{0,x}\partial_x^{-2}v_{0,yy}+\frac{15}{2}v_{0}^2v_{0,x},
\end{eqnarray}
where $y=t_{02}$ and $t=t_{05}$.

$\bullet$ The integrable equation corresponding to the case of $m=2$ and $n=6$ in (\ref{eq:KPhe})
can be derived from the following equations:
\begin{eqnarray}\label{eq:kp4-1}
&&\frac{\partial v_0}{\partial t_{06}}=\{v_0, H_0, H_6\},
\ \ \frac{\partial v_0}{\partial t_{02}}=\{v_0, H_0, H_2\},
\ \ \frac{\partial v_1}{\partial t_{02}}=\{v_1, H_0, H_2\},\nonumber\\
&&\frac{\partial v_2}{\partial t_{02}}=\{v_2, H_0, H_2\},
\ \ \frac{\partial v_3}{\partial t_{02}}=\{v_3, H_0, H_2\},
\ \ \frac{\partial v_4}{\partial t_{02}}=\{v_4, H_0, H_2\}.
\end{eqnarray}
The integrable equation is
\begin{eqnarray}\label{eq:KPE4-1}
\frac{\partial v_0}{\partial t}\!\!\!\!&=&\!\!\!\!3(v_{0,y})^2+9 v_0v_{0,x}\partial_x^{-1}v_{0,y}+6\partial_x^{-1}(v_{0,x}v_{0,yy})+12v_0^2v_{0,y}
+3v_{0,x}\partial_x^{-1}(v_0v_{0,y})\nonumber\\
&&\!\!\!\!-\frac{21}{4}\partial_x^{-1}(v_0v_{0,xxxy})+\frac{3}{4}
\partial_x^{-2}(v_0v_{0,yyy})+\frac{15}{4}\partial_x^{-2}(v_{0,yy})v_{0,y}
+\frac{15}{4}v_0\partial_{x}^{-2}(v_{0,yyy})
\nonumber\\
&&\!\!\!\!
+6\partial_x^{-1}(v_{0,xxy}\partial_x^{-1}v_{0,y})+6\partial_x^{-1}(v_{0,xx}
\partial_x^{-1}v_{0,yy})-\frac{3}{2}\partial_x^{-1}(v_0\partial_x^{-1}v_{0,yyy})
-\frac{3}{4}v_{0,x}v_{0,xy}\nonumber\\
&&\!\!\!\!
+\frac{33}{4}v_{0,xx}v_{0,y}+\frac{33}{4}v_0v_{0,xxy}-6v_{0,xy}
\partial_x^{-1}v_{0,y}+\frac{3}{4}v_{0,xxx}\partial_{x}^{-1}v_{0,y}
-6v_{0,x}\partial_x^{-1}v_{0,yy}\nonumber\\
&&\!\!\!\!+\frac{3}{4}v_{0,x}\partial_x^{-3}v_{0,yyy}
+\frac{9}{4}\partial_x^{-2}(v_{0y}v_{0,yy})-\frac{3}{2}\partial_x^{-1}(v_{0,y}\partial_x^{-1}v_{0,yy})
+\frac{3}{16}\partial_x^{-4}(v_{0,yyyyy})\nonumber\\
&&\!\!\!\!+\frac{5}{8}v_{0,yyy}-\frac{21}{4}\partial_x^{-1}(v_{0,y}v_{0,xxx})
-\frac{9}{4}\partial_x^{-1}(v_{0,x}\partial_x^{-2}v_{0,yyy})+\frac{3}{16}v_{0,xxxxy}\nonumber\\
&&\!\!\!\!
-\frac{9}{4}\partial_x^{-1}(v_{0,xy}\partial_x^{-2}v_{0,yy})
+\frac{9}{4}(\partial_x^{-1}v_{0,yy})(\partial_x^{-1}v_{0,y}),
\end{eqnarray}
where $y=t_{02}$ and $t=t_{06}$.

%%%%%%%%%%%%%%%%%%%%%%%%%%%%%%%%%%%%%%%%%%%%%%%%%%%%%%%%%%%%%%%%%%%%%%%%%
\section{ KP equation, KdV equation and $W_{1+\infty}$ 3-algebra }
%%%%%%%%%%%%%%%%%%%%%%%%%%%%%%%%%%%%%%%%%%%%%%%%%%%%%%%%%%%%%%%%%%%%%%%%%
One has already known that the $W_{1+\infty}$ algebra  is related to the
KP hierarchy.
Since the $W_{1+\infty}$ 3-algebra with a fixed generator $W^0_0$ in the
operator Nambu 3-bracket (\ref{eq:3-alg1}) recovers  the $W_{1+\infty}$ algebra,
it is natural to derive the  KP hierarchy from the $W_{1+\infty}$ 3-algebra.
To explore the deep connection between the $W_{1+\infty}$ 3-algebra and the
integrable equations, we need to deal with new case studies.
Hereafter we focus on the general case of the $W_{1+\infty}$ 3-algebra.
Due to the N-P evolution equation involving
two Hamiltonians, this may hide some still unknown relations between these Hamiltonians.
Thus it would be quite interesting to understand these Hamiltonians from the N-P evolution equation.

To start with let us assign the weights as follows:
\begin{eqnarray}\label{eq:weight}
 [x]=-1, [\partial_x]=1, [v_i]=i+2, [\partial t_{mn}]=m+n.
 \end{eqnarray}

{\bf Theorem }\quad For any Hamiltonian pair $(H_m, H_n)$ of the KP hierarchy,  $0\leq m<n$,
from the N-P evolution equations
\begin{subequations}\label{eq:KPn}
\begin{align}
\frac{\partial v_0}{\partial t_{mn}}=\{v_0, H_m, H_n\}, \label{eq:KPna}\\
\frac{\partial v_i}{\partial t_{02}}=\{v_i, H_0,  H_2\},\ \ \label{eq:KPnb}
\end{align}
\end{subequations}
where $i=0,1,\cdots, s-2$ and $s=m+n\geq 1$, it gives
the evolution equation with respect to $v_0$. The weight of equation is $s+2$.

{\bf Proof}\quad
By weight counting, it is easy to see that the pseudo-differential
operator $L$ (\ref{eq:L}) is homogeneous of weight 1,
from which we conclude that the weight of $L^{n+1}$ is $[L^{n+1}]=n+1$.
For any Hamiltonian $H_n$ (\ref{eq:KPGHamiltonian}), due to the weight $[Res L^{n+1}]=n+2$,
it implies that there are only finite $v_i$, $i=0,1,2,\cdots, n$ being permitted to emerge in $H_n$.
Note that for these  finite $v_i$,  $v_n$ has the highest weight $n+2$. Thus to determine
how many $v_i$ appearing in (\ref{eq:KPna}),
let us consider $\int\int\{v_0(x),v_m(y),v_n(z)\}dydz$.
By a direct calculation, we have
\begin{eqnarray}\label{eq:vmvn}
\int\int\{v_0(x),v_m(y),v_n(z)\}dydz=\sum_{p=1}^n C_n^p \partial_x^{p} v_{s-p}-\sum_{p=1}^m C_m^p \partial_x^{p} v_{s-p}.
\end{eqnarray}
In terms of (\ref{eq:vmvn}), we find that there are only finite $v_i$, $i=0,1,2,\cdots, s-1$ being permitted to emerge in
the evolution equation of $v_0$ (\ref{eq:KPna}).

After a straightforward calculation, (\ref{eq:KPnb}) gives its final form
\begin{eqnarray}\label{eq:viy}
 \frac{\partial v_i}{\partial t_{02}}&=&\{v_i,H_0,H_2\}
=2v_{i+1,x}+v_{i,xx}+2v_iv_0\nonumber\\
&-&2\sum_{p=0}^i(-1)^p C_i^p v_{i-p}\partial_x^pv_0,\quad i=0,1,\cdots,s-2.
\end{eqnarray}
Thus we have a recursion relation from (\ref{eq:viy})
\begin{eqnarray}\label{eq:recursion}
v_{i+1}(x)=\frac{1}{2}\partial_x^{-1}(v_{i,t_{02}}-v_{i,xx}-2v_iv_0+2\sum_{p=0}^i(-1)^p C_i^pv_{i-p}\partial_x^pv_0).
\end{eqnarray}
By means of the recursion relation (\ref{eq:recursion}), we may express $v_i$, $i=1,2,\cdots, s-1$ as
the function of $v_0$. Then substituting these $v_i$ into (\ref{eq:KPna}),
we obtain the evolution equation with respect to $v_0$.
Since the Nambu 3-bracket relation (\ref{eq:KP3-alg2}) and
the Hamiltonian (\ref{eq:KPGHamiltonian}) are homogeneous of weight,
in terms of (\ref{eq:vmvn}), it immediately gives the weight $s+2$ of the evolution equation with respect to $v_0$.
\EN

According to above theorem, we know that there are $\frac{m+n+1}{2}$ different Hamiltonian pairs
$(H_{\frac{m+n-1}{2}-i},$ $H_{\frac{m+n+1}{2}+i})$, $i=0,1,\cdots \frac{m+n-1}{2}$ such that (\ref{eq:KPn}) gives
the evolution equations of $v_0$ with the same weight $m+n+2$.

Now let us turn our attention to the KP equation.
Taking the Hamiltonian pairs $(H_m, H_n)$ to be $(H_0, H_3)$ and $(H_1, H_2)$, respectively,
we may derive the evolution equations of $v_0$ with the same weight from (\ref{eq:KPn}).
For the case of $(H_0, H_3)$,  we have derived the KP equation from (\ref{eq:KP1}).
Let us replace the Hamiltonian pair $(H_0, H_3)$ in (\ref{eq:KP1a}) by $(H_1, H_2)$,
thus we have
\begin{eqnarray}\label{eq:H1H2}
&&\frac{\partial v_0}{\partial t_{12}}=\{v_0, H_1, H_2\}=v_{2,x}+v_{1,xx}-6v_0v_{0,x}.
\end{eqnarray}
Substituting (\ref{eq:kpv1-4}) into (\ref{eq:H1H2}),
we  obtain
\begin{eqnarray}\label{eq:H1H2KP}
v_{0,yy}=(-4v_{0,t}+v_{0,xxx}+28v_0v_{0,x})_x,
\end{eqnarray}
where $y=t_{02}$ and $t=t_{12}$.
Under the scaling transformations
$y\rightarrow\frac{1}{\sqrt{-3}}y$ and $v_0\rightarrow\frac{3}{7}v_0$,
(\ref{eq:H1H2KP}) becomes  the usual KP equation (\ref{eq:KPE}).
What is surprising here is that we derive the KP equation from the N-P evolution
equation with the different Hamiltonian pairs.
This is an intriguing nontrivial result, since there is not such kind of
remarkable property in the framework of Poisson evolution equation.
It is worth to emphasize that the $W_{1+\infty}$ 3-algebra plays a
pivotal role in deriving the KP equation here. Due to the N-P evolution equation involving
two Hamiltonians, our analysis indicates that there is the deep relationship between
the Hamiltonians of KP hierarchy.

Comparing (\ref{eq:H1H2}) with (\ref{eq:KP1a}), we note
that the coefficients of fields $v_{1}$ and $v_{2}$ in (\ref{eq:KP1a})
are three times that of (\ref{eq:H1H2}).
To explore more relationship between the Hamiltonian pairs $(H_0, H_3)$ and $(H_1, H_2)$,
we consider the following evolution equation:
\begin{eqnarray}\label{eq:mH1H2}
&&\frac{\partial v_0}{\partial t_{03}}-3\frac{\partial v_0}{\partial t_{12}}
=\{v_0, H_0,  H_3\}-3\{v_0, H_1, H_2\}.
\end{eqnarray}
Substituting the expressions of (\ref{eq:KP1a}) and
(\ref{eq:H1H2}) into (\ref{eq:mH1H2}),
we obtain
\begin{eqnarray}\label{eq:lKdV}
&&-2\frac{\partial v_0}{\partial t}=v_{0,xxx}+24v_{0}v_{0,x},
\end{eqnarray}
where $t=t_{03}=t_{12}$.
Under the scaling transformations
$t\rightarrow-2t$ and $v_0\rightarrow\frac{1}{4}v_0$,
(\ref{eq:lKdV}) becomes  the well-known KdV equation
\begin{eqnarray}\label{eq:KdV}
&&\frac{\partial v_0}{\partial t}=v_{0,xxx}+6v_{0}v_{0,x}.
\end{eqnarray}

The connection between the KdV equation and the Virasoro algebra was first pointed out by
Gervais and Neveu \cite{Gervais1, Gervais2}.
The KdV equation is a bi-hamiltonian system.  It makes contact with the Virasoro algebra via the
second hamiltonian structure.
We have derived the  KP equation from the N-P evolution equation
with the different Hamiltonian pairs.
It is well-known that the KP hierarchy is introduced
as a generalization of KdV hierarchies.
Under the low dimensional reduction, the  KP equation reduces to the KdV equation.
As one of intriguing results in this paper, we find that the KdV equation
may be obtained from the N-P evolution equation  (\ref{eq:mH1H2})
without taking the low dimensional reduction.
More precisely, we may directly  derive the KdV equation from (\ref{eq:mH1H2})
by means of $W_{1+\infty}$ 3-algebra, where the Hamiltonians $H_i, i=0,1,2,3$
are the Hamiltonians of KP hierarchy.
It should also be stressed that without the condition of low dimensional reduction,
this result can not be achieved from the Poisson evolution equation
with the Hamiltonians of KP hierarchy. Therefore our investigation provides
new insight into the KdV equation.

Let us turn to consider the case of the Hamiltonian pairs $(H_m, H_n)$ with $m+n=4$.
For the case of $(H_0, H_4)$,  we have derived the integrable equation (\ref{eq:ekp2})
from (\ref{eq:KP2}).
If the first equation in (\ref{eq:KP2}) is replaced by
\begin{eqnarray}\label{eq:v0H1H3KP}
\frac{\partial v_0}{\partial t_{13}}&=&\{v_0,H_1,H_3\}
=2 v_{3,x}+3 v_{2,xx}+ v_{1,xxx}\nonumber\\
&&-6 v_1v_{0,x}-6 v_0v_{1,x}-3 v_0v_{0,xx}-3 (v_{0,x})^2,
\end{eqnarray}
then from (\ref{eq:v0H1H3KP}) and the last three equations in (\ref{eq:KP2}),
we obtain
\begin{eqnarray}\label{eq:H1H3kp2}
\frac{\partial^3 v_0}{\partial y\partial x^2}+4\frac{\partial v_0}{\partial t}
-\partial_x^{-2}\frac{\partial^3 v_0}{\partial y^3}+20v_{0,x}\partial_x^{-1}
\frac{\partial v_0}{\partial y}+16v_0\frac{\partial v_0}{\partial y}=0,
\end{eqnarray}
where $y=t_{02}$ and $t=t_{13}$.
Its single soliton solution is
\begin{equation}
v_0=\frac{k^2}{12}sech^2\frac{\xi}{2},
\end{equation}
where
\begin{eqnarray}\label{eq:xi}
\xi=k(\omega t+x+y)+c,
\end{eqnarray}
in which  $\omega=\frac{1}{4}(-k^2+1)$,  $k$ and $c$ are the constants.
Although (\ref{eq:H1H3kp2}) has the single soliton solution,
by performing  a Painlev\'{e} analysis of this equation, we note that it does not pass the test.
Hence (\ref{eq:H1H3kp2}) does not have the Painlev\'{e} integrable property.
In spite of this negative result it is instructive to
pursue the analysis of (\ref{eq:H1H3kp2}).
Taking the reduction $x=y$,
it is interesting to note that (\ref{eq:H1H3kp2}) may reduce to the KdV equation
(\ref{eq:KdV}) with the appropriate scaling transformations.

Under the low dimensional reduction,
we may also derive the KdV equation from the following
 N-P evolution equations with respect to $v_0$:
\begin{subequations}\label{eq:RKdV}
\begin{align}
\frac{\partial v_0}{\partial t_{04}}-2
\frac{\partial v_0}{\partial t_{13}}
=\{v_0, H_0, H_4\}-2\{v_0, H_1, H_3\}\ \ \ \nonumber\\
=2v_{1,xxx}+v_{0,xxxx}+12(v_{0,x})^2\ \ \ \ \nonumber\\
+12v_0v_{0,xx}+24v_0v_{1,x}+24v_1v_{0,x},\label{eq:RKdVa}\\
\frac{\partial v_0}{\partial t_{02}}=\{v_0, H_0, H_2\}=2 v_{1,x}+ v_{0,xx}.\ \ \ \ \ \ \ \ \ \ \ \ \ \ \ \label{eq:RKdVb}
\end{align}
\end{subequations}
Comparing  (\ref{eq:RKdVa}) with (\ref{eq:KP2a}), we observe
that there are not the fields $v_2$ and $v_3$ in (\ref{eq:RKdVa}).
Thus not as the case of (\ref{eq:KP2}), we do not need the evolution
equations of $v_1$ and $v_2$ here.
Eliminating $v_1$ via the evolution equations of $v_0$ (\ref{eq:RKdVb}), we can
rewrite (\ref{eq:RKdVa}) as
\begin{eqnarray}\label{eq:RRKdV}
-\frac{\partial v_0}{\partial t}=
\frac{\partial^3 v_0}{\partial y\partial x^2}
+12v_0\frac{\partial v_0}{\partial y}
+12v_{0,x}(\partial_x^{-1}\frac{\partial v_0}{\partial y}),
\end{eqnarray}
where $t=t_{04}=t_{13}$ and $y=t_{02}$.
Its single soliton solution is
\begin{eqnarray}
v_0=\frac{k^2}{8}sech^2\frac{\xi}{2},
\end{eqnarray}
where $\xi=k(-k^2t+x+y)+c$, $k$ and $c$ are the constants.
By performing  a Painlev\'{e} analysis of this equation, we note that it does not pass the test.
Taking the reduction $x=y$ and the scaling transformations
$t\rightarrow-t$ and $v_0\rightarrow\frac{1}{4}v_0$,
(\ref{eq:RRKdV}) reduces to the KdV equation (\ref{eq:KdV}).

%%%%%%%%%%%%%%%%%%%%%%%%%%%%%%%%%%%%%%%%%%%%%%%%%%%%%%%%%%%%%%%%%%%%%%%%%%%%%%%%%%%%%
\section{$W_{1+\infty}$ 3-algebra  and dispersionless KdV hierarchy}
%%%%%%%%%%%%%%%%%%%%%%%%%%%%%%%%%%%%%%%%%%%%%%%%%%%%%%%%%%%%%%%%%%%%%%%%%%%%%%%%%%%%%

In the previous section, we derived the (2+1)-dimensional equation (\ref{eq:H1H3kp2})
by means of the $W_{1+\infty}$ 3-algebra.
Even though this equation is not integrable,
nevertheless it still admits a single soliton solution and
reduces to the KdV equation under the low dimensional reduction.
In order to achieve a better understanding of the relation between $W_{1+\infty}$ 3-algebra and
(1+1)-dimensional integrable equations, we have to proceed to the detailed analysis.
Based on the theorem in the previous section, let us first give several examples of reduction.

$\bullet$ Hamiltonian pairs $(H_m, H_n)$ with $m+n=5$.

For this case, substituting the Hamiltonian pairs $(H_0, H_5)$, $(H_1, H_4)$ and $(H_2, H_3)$ into
(\ref{eq:KPn}), respectively,  we may derive the evolution equations of $v_0$ with the same weight $7$.
We have given the integrable equation (\ref{eq:ekp3}) from (\ref{eq:KP3})
with the Hamiltonian pair $(H_0, H_5)$.
Let us now replace the first equation in (\ref{eq:KP3})  by
\begin{eqnarray}\label{eq:H1H4KP}
\frac{\partial v_0}{\partial t_{14}}&=&\{v_0, H_1, H_4\}\nonumber\\
&=&3 v_{4,x}+6 v_{3,xx}+4 v_{2,xxx}+ v_{1,xxxx}-4 v_2v_{0,x}-4 v_0v_{2,x}
-4 v_1v_{1,x}\nonumber\\
&&-6 v_{0,x}v_{1,x}-4 v_0v_{1,xx}-2 v_1v_{0,xx}-12 v_{0,x}v_{0,xx}
-6v_0v_{0,xxx}-30 v_0^2v_{0,x},
\end{eqnarray}
then from (\ref{eq:H1H4KP}) and the last four equations in (\ref{eq:KP3}),
we obtain
\begin{eqnarray}\label{eq:kp-3-1}
\frac{\partial v_0}{\partial t}&=&\frac{3}{16}\partial_x^{-3}v_{0,yyyy}-\frac{1}{8}v_{0,xyy}-\frac{5}{2}v_{0,y}\partial_x^{-1}v_{0,y}
-\frac{17}{2}v_{0,x}v_{0,xx}\nonumber\\
&&-\frac{21}{4}v_{0}v_{0,xxx}-\frac{1}{16}v_{0,xxxxx}
-\frac{5}{2}v_{0}\partial_x^{-1}v_{0,yy}+\frac{3}{4}\partial_x^{-1}(v_{0,y})^2\nonumber\\
&&+\frac{3}{4}\partial_x^{-1}(v_0v_{0,yy})
-\frac{13}{4}v_{0,x}\partial_x^{-2}v_{0,yy}-\frac{39}{2}v_{0}^2v_{0,x},
\end{eqnarray}
where $y=t_{02}$ and $t=t_{14}$.
Its single soliton solution is
\begin{equation}
v_0 = \frac{1}{78}\frac{(3790-110\sqrt{1249})k^2+531-15\sqrt{1249}-(11370-330\sqrt{1249})k^2sech^2\xi}{-17+\sqrt{1249}},
\end{equation}
where $\xi$ is given by (\ref{eq:xi}) with
$\omega=-\frac{1}{624}\frac{(141441304\sqrt{1249}-4998879128)k^4+134406117-3805749\sqrt{1249}}{-262507+7451\sqrt{1249}}$.
To test the integrability of this equation, we perform the
Painlev\'{e} analysis of (\ref{eq:kp-3-1}). Unfortunately it does not pass the test.

If the first equation in (\ref{eq:KP3}) is replaced by
\begin{eqnarray}\label{eq:kp-023}
\frac{\partial v_0}{\partial t_{23}}&=&\{v_0, H_2, H_3\}\nonumber\\
&=& v_{4,x}+2 v_{3,xx}+ v_{2,xxx}+3 v_2v_{0,x}+3 v_0v_{2,x}-18 v_1v_{1,x}\nonumber\\
&&-6 v_{0,x}v_{1,x}+3 v_0v_{1,xx}-9 v_1v_{0,xx}-3 v_{0,x}v_{0,xx}+3v_0v_{0,xxx}+24 v_0^2v_{0,x},
\end{eqnarray}
we can derive
the following equation:
\begin{eqnarray}\label{eq:kp-3-2}
\frac{\partial v_0}{\partial t}&=&\frac{1}{16}\partial_x^{-3}v_{0,yyyy}-\frac{1}{8}v_{0,xyy}-5v_{0,y}\partial_x^{-1}v_{0,y}
+\frac{9}{4}v_{0,x}v_{0,xx}+\frac{5}{2}v_{0}v_{0,xxx}\nonumber\\
&&+\frac{1}{16}v_{0,xxxxx} +\frac{1}{4}v_{0}\partial_x^{-1}v_{0,yy}+\frac{1}{4}\partial_x^{-1}(v_{0,y})^2+\frac{1}{4}\partial_x^{-1}(v_0v_{0,yy})
+21v_{0}^2v_{0,x},
\end{eqnarray}
where $y=t_{02}$ and $t=t_{23}$.
By applying the Painlev\'{e} analysis to (\ref{eq:kp-3-2}), we find that it does not pass the test.
As the case of (\ref{eq:kp-3-1}), (\ref{eq:kp-3-2}) has
the single soliton solution as follows:
\begin{equation}
v_0 =\frac{31}{140}-\frac{5}{14}k^2+\frac{15}{14}k^2sech^2\xi,
\end{equation}
where $\xi$ is given by (\ref{eq:xi}) with
$\omega =\frac{3}{28}k^4+\frac{67}{700}$.

Under the reduction $y=0$,  (\ref{eq:ekp3}), (\ref{eq:kp-3-1}) and (\ref{eq:kp-3-2}) reduce to
 the following  equations:
\begin{eqnarray}\label{eq:KdV1}
&&\frac{\partial v_0}{\partial t}=\frac{1}{16}v_{0,xxxxx}+
\frac{5}{4}v_0v_{0,xxx}+
\frac{5}{2}v_{0,x}v_{0,xx}+\frac{15}{2}v_0^2v_{0,x},
\end{eqnarray}
\begin{eqnarray}\label{eq:KdV2}
&&\frac{\partial v_0}{\partial t}=-\frac{1}{16}v_{0,xxxxx}
-\frac{21}{4}v_0v_{0,xxx}
-\frac{17}{2}v_{0,x}v_{0,xx}-\frac{39}{2}v_0^2v_{0,x}
\end{eqnarray}
and
\begin{eqnarray}\label{eq:KdV3}
&&\frac{\partial v_0}{\partial t}=\frac{1}{16}v_{0,xxxxx}
+\frac{5}{2}v_0v_{0,xxx}
+\frac{9}{4}v_{0,x}v_{0,xx}+21v_0^2v_{0,x},
\end{eqnarray}
respectively.
In the dispersionless limit
$\partial_t\rightarrow \epsilon \partial_t$ and $\partial_x\rightarrow \epsilon \partial_x$ with $\epsilon\rightarrow 0$,
(\ref{eq:KdV1}), (\ref{eq:KdV2}) and (\ref{eq:KdV3}) reduce to the following dispersionless KdV equation with the appropriate scaling transformations:
\begin{eqnarray}
v_{0,t}=\frac{15}{8}v_0^2v_{0,x}.
\end{eqnarray}
Note that (\ref{eq:KdV1}) is the completely integrable Lax's fifth order KdV equation \cite{Lax},
but (\ref{eq:KdV2}) and (\ref{eq:KdV3}) are not integrable.
The fifth order KdV equations describe motions of long waves in shallow water under gravity and in a one-dimensional
nonlinear lattice and has wide applications in quantum mechanics and nonlinear optics.
Thus a great deal of research work has been invested  for the study of the fifth order KdV equations.
Recently the following general fifth order KdV equation has been investigated \cite{Salas}:
\begin{eqnarray}\label{eq:GFKdV}
\frac{\partial v}{\partial t}=\omega v_{xxxxx}+\alpha vv_{xxx}
+\beta v_{x}v_{xx}+\gamma v^2v_{x},
\end{eqnarray}
where $\alpha$, $\beta$, $\gamma$ and $\omega$ are the arbitrary real parameters.
By using the exp function method, the  solutions of (\ref{eq:GFKdV}) have also
been presented.
It can be very easily seen that (\ref{eq:KdV2}) and (\ref{eq:KdV3})
are the special cases  of (\ref{eq:GFKdV}).

$\bullet$ Hamiltonian pairs $(H_m, H_n)$ with $m+n=7$.

The integrable equation corresponding to the case of $m=2$ and $n=7$ in (\ref{eq:KPhe})
can be derived from the following equations:
\begin{eqnarray}\label{eq:kp07}
&&\frac{\partial v_0}{\partial t}=\{v_0, H_0, H_7\},
\ \ \frac{\partial v_0}{\partial y}=\{v_0, H_0, H_2\},
\ \ \frac{\partial v_1}{\partial y}=\{v_1, H_0, H_2\},\nonumber\\
&&\frac{\partial v_2}{\partial y}=\{v_2, H_0, H_2\},
\ \ \frac{\partial v_3}{\partial y}=\{v_3, H_0, H_2\},
\ \ \frac{\partial v_4}{\partial y}=\{v_4, H_0, H_2\},\nonumber\\
&&\frac{\partial v_5}{\partial y}=\{v_5, H_0, H_2\}.
\end{eqnarray}
Due to too many terms in this integrable equation, we do not exhibit its explicit expression here.
Let us replace
the first equation in (\ref{eq:kp07})  by
\begin{eqnarray}
&&\frac{\partial v_0}{\partial t}=\{v_0, H_1, H_6\},\nonumber\\
&&\frac{\partial v_0}{\partial t}=\{v_0, H_2, H_5\},\nonumber\\
&&\frac{\partial v_0}{\partial t}=\{v_0, H_3, H_4\},
\end{eqnarray}
respectively.
We may also derive three non-integrable equations.
Taking the reduction $y=0$ and  the dispersionless limit,
these four equations  reduce to the following dispersionless KdV
equation with the appropriate scaling transformations:
\begin{eqnarray}
v_{0,t}=\frac{35}{16}v_0^3v_{0,x}.
\end{eqnarray}

The dispersionless KdV hierarchy is a well studied
integrable system \cite{Das, Brunelli}. It is given by
\begin{eqnarray}\label{eq:DKdVH}
\frac{\partial u(t, x) }{\partial
 t}&=&A_{n-1}u^{n-1}u_{x}\nonumber\\
&=&\left\{\begin{array}{ccccccccc}
 u_{x}, &n=1,\\
\\
\frac{3}{2} u u_{x}, &n=2,\\
\\
\frac{15}{8} u^2 u_{x}, &n=3,\\
\\
\frac{35}{16} u^3 u_{x}, &n=4,\\
\\
\vdots &, \\
\end{array}\right.
\end{eqnarray}
where $A_{n}$ is given by
\begin{eqnarray}\label{eq:An}
A_{n}=\frac {(2n+1)!!}{(2n)!!}.
\end{eqnarray}

We have derived the third and fourth order dispersionless KdV equations
from the N-P evolution equation.
For the second order dispersionless KdV equation in (\ref{eq:DKdVH}),
it is easy to obtain
from  (\ref{eq:KPE}) and (\ref{eq:H1H2KP}) under the low
dimensional reduction with the dispersionless limit, respectively.
From above several examples, it is of interest to note that for a higher-order
dispersionless KdV equation, we may derive it from the N-P evolution
equation with the different Hamiltonian pairs.
It indicates that there is the intrinsic equivalent relation between the
Hamiltonian pairs. We call this intrinsic relation between the
Hamiltonian pairs "degeneration".
Based on above analysis, we present a conjecture as follows:

{\bf Conjecture }\quad For any Hamiltonian pair $(H_m,H_n)$ of the KP hierarchy, $0\leq m<n$,
when $s = m+ n$ is odd,
we may choose $\frac{s+1}{2}$ different Hamiltonian pairs $(H_m, H_n)$
for (\ref{eq:KPna}) such that  under the low
dimensional reduction with the dispersionless limit,
(\ref{eq:KPn}) leads to the dispersionless KdV hierarchy.

Let us make a comment about this conjecture.
When $s=1$, from (\ref{eq:KPna}), it is easy to obtain
\begin{eqnarray}
\frac{\partial v_0}{\partial t_{01}}=\{v_0, H_0,  H_1\}=v_{0,x}.
\end{eqnarray}
This is the first equation in (\ref{eq:DKdVH}).

When $s=m+n$ is odd and $s\ge 3$,
 let us choose the following $\frac{s+1}{2}$ different Hamiltonian pairs $(H_{\frac{s-1}{2}-i}, H_{\frac{s+1}{2}+i})$, $i=0,1,\cdots \frac{s-1}{2}$.
 Based on the theorem in the previous section, we can derive $\frac{s+1}{2}$ different evolution equations with respect to
 $v_0$ with odd weight $s+2$.  Under the low dimensional reduction $t_{02}=0$ and the dispersionless limitation, i.e.,
$\partial_{t_{mn}}\rightarrow \epsilon \partial_{t_{mn}}$ and $\partial_x\rightarrow \epsilon \partial_x$ with $\epsilon\rightarrow 0$,
all higher derivation terms of $v_0$ in the evolution equations are vanishing.  Thus we have the following equation with
the weight $s+2$:
\begin{eqnarray}\label{eq:alph}
\frac{\partial v_0}{\partial t_{mn}}=\alpha_{mn} v_0^{\frac{s-1}{2}}v_{0,x},
\end{eqnarray}
where $\alpha_{mn}$ is a constant.
When $\alpha_{mn}\neq 0$, by taking the appropriate scaling transformation, (\ref{eq:alph}) leads to the dispersionless KdV hierarchy. However, how to prove $\alpha_{mn}\neq 0$ is the most challenging problem.

%%%%%%%%%%%%%%%%%%%%%%%%%%%%%%%%%%%%%%%%%%%%%%%%%%%%%%%%%%%%%%%%%%%%%%%%%%%%%%%%%%%%%
\section{$W_{1+\infty}$ 3-algebra  and non-linear Schr\"{o}dinger equations}
%%%%%%%%%%%%%%%%%%%%%%%%%%%%%%%%%%%%%%%%%%%%%%%%%%%%%%%%%%%%%%%%%%%%%%%%%%%%%%%%%%%%%

%%%%%%%%%%%%%%%%%%%%%%%%%%%%%%%%%%%%%%%%%%%%%%%%%%%%%%%%%%%%%%%%%%%%%%%%%%%%%%%%%%%%
\subsection{Non-linear Schr\"{o}dinger equation}
%%%%%%%%%%%%%%%%%%%%%%%%%%%%%%%%%%%%%%%%%%%%%%%%%%%%%%%%%%%%%%%%%%%%%%%%%%%%%%%%%%%%%

The non-linear Schr\"{o}dinger equation is given by
\begin{equation}\label{eq:NLSE}
{\bf i}\frac{\pa\psi}{\pa t}+\frac{\pa^2\psi}{\pa x^2}+2|\psi|^2\psi=0,
\end{equation}
where $\psi$ is a complex bosonic field.

The non-linear Schr\"{o}dinger equation possesses an infinite set of conserved
quantities $\hat{H}_n$, given by the formula
\begin{equation}\label{eq:NLSconserved}
\hat{H}_n=-(-{\bf i})^n\int\psi\Gamma_n(\psi,\psi^*)dx,
\end{equation}
where
\begin{equation}
\Gamma_0=\psi^*,\quad\quad \Gamma_1=\Gamma_{0,x}, \quad\quad \Gamma_{n+1}=\Gamma_{n,x}+\psi\sum_{p=0}^{n-1}\Gamma_p\Gamma_{n-1-p}, \quad\quad n\geq 1.
\end{equation}
The first few such Hamiltonians are
\begin{eqnarray}
\hat{H}_0 &=& -\int\psi^*\psi dx, \nonumber\\
\hat{H}_1 &=& -{\bf i}\int\psi^*\psi_xdx,\nonumber\\
\hat{H}_2 &=& \int(\psi^*\psi_{xx}+\psi^{*2}\psi^2)dx,\nonumber\\
\hat{H}_3 &=& {\bf i}\int(\psi^*\psi_{xxx}+3\psi^{*2}\psi\psi_x)dx,\nonumber\\
\hat{H}_4 &=& -\int (\psi^*\psi_{xxxx}+4\psi^{*2}\psi\psi_{xx}+2\psi^{*2}\psi_x^2-2\psi^*\psi(\psi^*\psi_x)_x \nonumber\\
&&+\psi^*\psi(\psi^*\psi)_{xx}+2(\psi^*\psi)^3)dx,\nonumber\\
&&\vdots.
\end{eqnarray}

The non-linear Schr\"{o}dinger equation (\ref{eq:NLSE}) can be obtained from
\begin{eqnarray}\label{eq:psi-h2}
\frac{\pa\psi}{\pa t}&=&\{\psi(x), \hat{H}_2\},
\end{eqnarray}
where the Poisson
brackets of the complex bosonic  field are given by  \cite{Freeman}
\begin{equation}\label{eq:NLSPoisson}
\{\psi^*(x),\psi(y)\}=-{\bf i}\d(x-y), \quad\quad \{\psi(x),\psi(y)\}=\{\psi^*(x),\psi^*(y)\}=0.
\end{equation}

It is known that there is  a realization of $W_{1+\infty}$ algebra  in terms of
a complex bosonic field with Poisson brackets (\ref{eq:NLSPoisson}).
Taking
\begin{equation}\label{eq:relation}
v_r(x)={\bf i}(-1)^r\psi^*(x)\psi^{(r)}(x),
\end{equation}
where $\psi^{(r)}(x)=\frac{d^r\psi(x)}{dx^r}$,
by means of (\ref{eq:NLSPoisson}), it is easy to verify that the generators (\ref{eq:relation})
satisfy (\ref{eq:fwalgebra}).
Due to this kind of realization of $W_{1+\infty}$ algebra, the connection between
the non-linear Schr\"{o}dinger equation and the KP hierarchy has been established
by Freeman and West \cite{Freeman}.

For the case of $W_{1+\infty}$ 3-algebra (\ref{eq:KP3-alg2}), we find that its realization
can be achieved in terms of a complex bosonic field with Nambu 3-brackets given by
\begin{eqnarray}\label{eq:NLSPoisson33}
&&\{\psi^{(r)}(x), \psi^*(y)\psi^{(s)}(y), \psi^*(z)\}  = -{\bf i} \pa_x^r \delta(x-y) \pa_y^s\delta(y-z),\nonumber\\
&&\{\psi^*(x), \psi^*(y)\psi^{(s)}(y), \psi^*(z)\} = \{\psi^{(r)}(x), \psi^*(y)\psi^{(s)}(y), \psi^{(h)}(z)\} = 0.
\end{eqnarray}
The proof is straightforward. One can confirm the $W_{1+\infty}$ 3-algebra (\ref{eq:KP3-alg2}) by substituting
$v_r(x)=\frac{\sqrt{6}}{2}(1+{\bf i})(-1)^r\psi^*(x)\psi^{(r)}(x)$ and using the
Nambu 3-brackets (\ref{eq:NLSPoisson33}).

As the cases of (\ref{eq:vHv}) and (\ref{eq:HPR}),
in terms of the Nambu 3-bracket relations (\ref{eq:NLSPoisson33}), we have also the similar equivalent relations
\begin{equation}
\{\psi, \hat{H}_0, \psi^*\}\cong \{\psi, \psi^*\}, \quad\quad
\{\psi, \hat{H}_0, \hat{H}_n\}\cong  \{\psi, \hat{H}_n\}.
\end{equation}
Thus we may also derive the non-linear Schr\"{o}dinger equation from
the N-P evolution equation
\begin{equation}\label{eq:NLS-NP-evolution}
\frac{\pa\psi}{\pa t}=\{\psi, \hat{H}_0, \hat{H}_2\}={\bf i}(\frac{\pa^2\psi}{\pa x^2}+2|\psi|^2\psi).
\end{equation}

Based on the $W_{1+\infty}$ 3-algebra (\ref{eq:KP3-alg2}), we have derived the KP and KdV
equations from the N-P evolution equation with the first few  Hamiltonian pairs
of KP hierarchy. Note that the $W_{1+\infty}$ 3-algebra (\ref{eq:KP3-alg2}) can be
realized in terms of the complex bosonic field with the Nambu 3-bracket relations
(\ref{eq:NLSPoisson33}). Based on  (\ref{eq:NLSPoisson33}), let us focus on investigating
the N-P evolution equations with the first few Hamiltonian pairs of non-linear Schr\"{o}dinger equation.
A list of the corresponding nonlinear evolution equations is given as follows:

$\bullet$ Hamiltonian pair $(\hat{H}_0, \hat{H}_3)$
\begin{eqnarray}\label{eq:psi-H03}
\frac{\pa\psi}{\pa t} = \{\psi, \hat{H}_0, \hat{H}_3\}
= -(\frac{\pa^3\psi}{\pa x^3}+6|\psi|^2\frac{\pa\psi}{\pa x}),
\end{eqnarray}
This equation is the so-called integrable complex mKdV equation.

$\bullet$ Hamiltonian pair $(\hat{H}_1, \hat{H}_2)$
\begin{eqnarray}\label{eq:psi-H12}
\frac{\pa \psi}{\pa t} = \{\psi, \hat{H}_1, \hat{H}_2\}
=-(\frac{\pa^3\psi}{\pa x^3}+2\frac{\pa |\psi|^2\psi}{\pa x}).
\end{eqnarray}
Its single soliton solution is
$\psi=k_1{\rm sech} (k_1x-k_1^3t)e^{{\bf i}k_2}$,
where $k_1$ and $k_2$ are the real constants.

$\bullet$ Hamiltonian pair $(\hat{H}_1, \hat{H}_3)$
\begin{eqnarray}\label{eq:psi-H13}
\frac{\pa \psi}{\pa t} = \{\psi, \hat{H}_1, \hat{H}_3\}
= -{\bf i}(\frac{\pa^4\psi}{\pa x^4}+6\frac{\pa}{\pa x}(|\psi|^2\frac{\pa\psi}{\pa x})).
\end{eqnarray}
Its single soliton solution is
$\psi=k{\rm sech} (kx-\frac{8\sqrt{3}}{9}k^4t)e^{{\bf i}(-\frac{\sqrt{3}k}{3}x+\frac{8}{9}k^4t})$,
where $k$ is a real constant.

To test the integrability of (\ref{eq:psi-H12}) and (\ref{eq:psi-H13}),
we  perform the Painlev\'{e} analysis and find that they
does not have the Painlev\'{e} integrable property.

For the Hamiltonian pairs $(H_0,H_3)$ and $(H_1,H_2)$ of KP hierarchy,
we have found that there is the deep relationship between them.
For the case of the Hamiltonian pairs of non-linear Schr\"{o}dinger equation,
as done in that of case (\ref{eq:mH1H2}), let us consider the following evolution equation:
\begin{eqnarray}\label{combine12}
\frac{\pa\psi}{\pa t} = \{\psi, \hat{H}_0, \hat{H}_3\}+3\{\psi, \hat{H}_1, \hat{H}_2\}.
\end{eqnarray}
Substituting the expressions of (\ref{eq:psi-H03}) and
(\ref{eq:psi-H12}) into (\ref{combine12}),
we obtain
\begin{eqnarray}\label{combine}
\frac{\pa\psi}{\pa t}=
 -4\frac{\pa^3\psi}{\pa x^3}-12|\psi|^2\frac{\pa\psi}{\pa x}-6\psi\frac{\pa |\psi|^2}{\pa x}.
\end{eqnarray}
Under the scaling transformations $\psi\rightarrow\sqrt{2}\psi$ and $t\rightarrow\frac{1}{4}t$,
we may rewrite (\ref{combine}) as
\begin{equation}\label{eq:SSnls}
\frac{\pa\psi}{\pa t}+\frac{\pa^3\psi}{\pa x^3}+6|\psi|^2\frac{\pa\psi}{\pa x}+3\psi\frac{\pa |\psi|^2}{\pa x}=0.
\end{equation}
We immediately recognize that (\ref{eq:SSnls}) is nothing but the integrable Sasa-Satsuma equation \cite{Sasa}.

%%%%%%%%%%%%%%%%%%%%%%%%%%%%%%%%%%%%%%%%%%%%%%%%%%%%%%%%%%%%%%%%%%%%%%%%%%%%%%%%%%%%
\subsection{ An application in optical soliton}
%%%%%%%%%%%%%%%%%%%%%%%%%%%%%%%%%%%%%%%%%%%%%%%%%%%%%%%%%%%%%%%%%%%%%%%%%%%%%%%%%%%%%

It is well-known that the nonlinear Schr\"{o}dinger equation plays a critical role in
the study of optical solitons for optical fiber communications.
To describe a large number of nonlinear effects in optical fibers,
the effects of the higher-order terms on the nonlinear Schr\"{o}dinger
equation have been paid more attention.
The propagation of optical field in a monomode optical fiber may be described by the following
generalized nonlinear Schr\"{o}dinger equation \cite{Agrawal}:
\begin{equation}\label{eq:GNLSE}
\frac{\pa A}{\pa Z}+\frac{\alpha}{2}A+\frac{{\bf i}\beta_2}{2}\frac{\pa^2 A}{\pa T^2}-\frac{\beta_3}{6}\frac{\pa^3 A}{\pa T^3}={\bf i}\gamma(|A|^2A+\frac{{\bf i}}{\omega_0}\frac{\pa |A|^2A}{\pa T}-T_RA\frac{\pa |A|^2}{\pa T}),
\end{equation}
where $Z$ is the normalized propagation distance, $T$ is the normalized retarded time, i.e., time in the frame of reference moving with the wave packet, $A(Z,T)$ denotes the normalized slowly varying complex wave packet envelope in the electric filed, and $|A|^2$ represents  the optical power.
(\ref{eq:GNLSE}) is an important nonlinear equation for studying higher-order
nonlinear effects in optical fibers.
The second and third terms on the left side of (\ref{eq:GNLSE}) describe the effects of fiber
loss and chromatic dispersion, respectively.
The term proportional to $\beta_3$ governs the effects of third-order dispersion and
becomes important for ultrashort pulses because of their wide bandwidth.
The first term on the right side of (\ref{eq:GNLSE}) describes the nonlinear effects
of self-phase modulation.
The terms proportional to $\omega_0^{-1}$ and $T_R$ are
responsible for the phenomenon of self-steepening and
the Raman-induced frequency shift, respectively.

Based on the results in the previous subsection, let us turn to derive generalized nonlinear
Schr\"{o}dinger equation in the framework of Nambu mechanics.
For the Hamiltonian pairs $(\hat{H}_0, \hat{H}_2)$ and $(\hat{H}_1, \hat{H}_2)$,
we consider the following N-P evolution equation:
\begin{eqnarray}\label{eq:NLSE2}
\frac{\pa\psi}{\pa t}=-\frac{\beta_2}{2}\{\psi, \hat{H}_0, \hat{H}_2\}
-\frac{\beta_3}{6}\{\psi, \hat{H}_1, \hat{H}_2\},
\end{eqnarray}
where the parameters $\beta_2$ and $\beta_3$ in (\ref{eq:GNLSE}) are introduced into (\ref{eq:NLSE2}).

By means of (\ref{eq:NLS-NP-evolution}) and (\ref{eq:psi-H12}), (\ref{eq:NLSE2}) leads to
\begin{eqnarray}\label{eq:NLSE3}
\frac{\pa\psi}{\pa t}=
-\frac{{\bf i}\beta_2}{2}\frac{\pa^2\psi}{\pa x^2}+\frac{\beta_3}{6}\frac{\pa^3\psi}{\pa x^3}
-{\bf i}\beta_2|\psi|^2\psi+\frac{\beta_3}{3}\frac{\pa |\psi|^2\psi}{\pa x}.
\end{eqnarray}
Its single soliton solution is
\begin{equation}
\psi = k{\rm sech} (kx+\frac{\beta_3}{6}k^3t)e^{-\frac{1}{2}{\bf i}\beta_2k^2t},
\end{equation}
where $k$ is a real constant.

Comparing (\ref{eq:GNLSE}) with (\ref{eq:NLSE3}), we observe that
in absence of the fiber loss and Raman-induced frequency shift terms, i.e., $\alpha=T_R=0$,
(\ref{eq:GNLSE}) reduces to (\ref{eq:NLSE3}) with $\gamma=-\beta_2$, and $\omega_0=\frac{3\beta_2}{\beta_3}$.

We have already seen that the generalized nonlinear Schr\"{o}dinger equation can be derived
in terms of the Hamiltonian pairs $(\hat{H}_0, \hat{H}_2)$ and $(\hat{H}_1, \hat{H}_2)$.
It should be pointed out that we may also derive the similar result just from the N-P evolution
equation with the Hamiltonian pair $(\hat{H}_1, \hat{H}_2)$ under the appropriate variable transformations.
For the Hamiltonian pair $(\hat{H}_1, \hat{H}_2)$, the corresponding N-P evolution equation
is given by (\ref{eq:psi-H12}).
Let us take the variable transformations as follows:
\begin{eqnarray}\label{eq:vt}
t &=& \frac{\gamma\beta_3}{2\beta_2}\sqrt{\frac{-3\gamma}{\beta_2}}Z,\nonumber\\
x &=& \frac{1}{2\beta_3}\sqrt{\frac{-3\gamma}{\beta_2}}(2\beta_3T+\beta_2^2Z),\nonumber\\
A &=& \psi(t, x)\exp{{\bf i}B},\nonumber\\
B &=& \frac{\beta_2}{\beta_3}(T+\frac{\beta_2^2}{3\beta_3}Z)+\frac{\alpha}{2}{\bf i}Z.
\end{eqnarray}
It is not difficult to prove that under the variable transformations (\ref{eq:vt}),
(\ref{eq:psi-H12}) becomes (\ref{eq:GNLSE}) with $T_R=0$ and $\omega_0=\frac{\beta_2}{\beta_3}$.

%%%%%%%%%%%%%%%%%%%%%%%%%%%%%%%%%%%%%%%%%%%%%%%%%%%%%%%%%%%
\section{Concluding Remarks}
%%%%%%%%%%%%%%%%%%%%%%%%%%%%%%%%%%%%%%%%%%%%%%%%%%%%%%%%%%%

The KP hierarchy is an important integrable system which can be regarded as
the generalization of KdV hierarchies.
It is well-known that there is a remarkable connection between the integrable
systems and the infinite-dimensional conformal algebra and its extensions.
For the KP hierarchy, one has known that its first Hamiltonian structure
 is related to the $W_{1+\infty}$ algebra.
To discuss the relation between the infinite-dimensional 3-algebra and the KP
hierarchy, we constructed the $W_{1+\infty}$ 3-algebra which
does not satisfy the FI condition,
but the BI condition  holds for this ternary algebra.
By introducing  the FTF,  we rewrote the $W_{1+\infty}$ 3-algebra
as the Nambu 3-bracket structure of the FTF.
We noted that when there is a fixed generator $W^0_0$ in the operator Nambu 3-bracket,
the $W_{1+\infty}$ 3-algebra may reduce to the $W_{1+\infty}$ algebra.
Thus based on this special $W_{1+\infty}$ 3-algebra,
we derived the KP hierarchy from the N-P evolution equation with the Hamiltonian pairs $(H_0,H_i)$.

To explore the relation between the general case of the  $W_{1+\infty}$ 3-algebra and
integrable equations,  we replaced the Hamiltonian pair $(H_0,H_3)$ by $(H_1,H_2)$ in
the N-P evolution equation  (\ref{eq:KP1}).
An intriguing feature is that it still leads to the KP equation.
Moreover without taking the low dimensional reduction,
we directly derived the KdV equation by means of the N-P evolution equation
with the Hamiltonian pairs $(H_0,H_3)$ and $(H_1,H_2)$. It did provide new insight into the KP and
KdV equations. Furthermore we investigated the N-P evolution equation (\ref{eq:KPn})
with the different Hamiltonian pairs and established the connections between the
$W_{1+\infty}$ 3-algebra and the dispersionless KdV equations.
We pointed out that there is an intrinsic equivalent relation between the
Hamiltonian pairs of KP hierarchy, i.e.,  "degeneration".
We also presented a conjecture with respect to the  "degeneration".
For the Hamiltonians of KP hierarchy, one did only know that these Hamiltonians  are in involution.
Our investigation turned out that the intrinsic relationship between these Hamiltonians
is actually more complicated than has been previously recognized.

For the $W_{1+\infty}$ 3-algebra, we also presented its realization in terms of a
complex bosonic field with Nambu 3-brackets (\ref{eq:NLSPoisson33}).
Based on (\ref{eq:NLSPoisson33}), we derived the (generalized) nonlinear Schr\"{o}dinger
equation, complex mKdV equation and Sasa-Satsuma equation
in the framework of Nambu mechanics. Moreover an application
in optical soliton has been given for the generalized
nonlinear Schr\"{o}dinger equation derived from N-P evolution equation.
It is worthwhile to mention the (2+1)-dimensional non-integrable equations
presented in this paper, which are derived from N-P evolution equation
with the Hamiltonian pairs of KP hierarchy and have the single soliton solutions.
Since the (higher-order) KP and KdV equations have been widely used in various fields
such as fluid physics, plasma physics and quantum field theory,
the applications of the (2+1)-dimensional evolution equations mentioned above
might be of interest in these fields.

Although the $W_{1+\infty}$ 3-algebra does not satisfy the so-called FI condition,
our analysis suggests that there still exist the much
deeper connections between the $W_{1+\infty}$ 3-algebra and the
integrable equations. More properties with respect to their relations  still deserve further study.
We believe that the infinite-dimensional 3-algebras may shed new light on the integrable systems.

%%%%%%%%%%%%%%%%%%%%%%%%%%%%%%%%%%%%%%%%%%%%%%%%%%%%%%%%%%%
\section*{Acknowledgements}
%%%
%%%%%%%%%%%%%%%%%%%%%%%%%%%%%%%%%%%%%%%%%%%%%%%%%%%%
The authors are indebted to Profs. S.Y. Lou and Q.P. Liu, Drs. D.S. Wang, J.P.
Cheng and M.X. Zhang for their valuable discussions.
The authors are grateful to Morningside Center of Chinese Academy of
Sciences for providing excellent research environment and financial support
to our seminar in mathematical physics.
This work is partially supported by NSF projects
(11375119, 11031005 and 11171329), KZ201210028032.

%%%%%%%%%%%%%%%%%%%%%%%%%%%%%%%%%%%%%%%%%%%%%%%%%%%%%%%%%%%%%%%%%%%%%%%%%%%%%%%

\end{document}